\documentclass[aps,preprint]{revtex4}

\usepackage{amssymb}
\usepackage{amsmath}
\usepackage{epsfig}
\usepackage{epstopdf}
\usepackage{bm}
\usepackage{graphicx,epsfig}
\usepackage{mathrsfs}
\usepackage{dcolumn}
\usepackage{color}
\usepackage{natbib}
\usepackage{CJK}
\def\be{\begin{equation}}
\def\ee{\end{equation}}
\def\bea{\begin{eqnarray}}
\def\eea{\end{eqnarray}}

\allowdisplaybreaks[2]

\begin{document}

\title{Topological Dissipative Photonics and Topological Insulator Lasers in Synthetic Time-Frequency Dimensions}
\author{Zhaohui Dong$^{1}$, Xianfeng Chen$^{1,2,3}$, Avik Dutt$^{4,\dagger }$, and Luqi Yuan$^{1,*}$}
\affiliation{$^1$State Key Laboratory of Advanced Optical Communication Systems and Networks, School of Physics and Astronomy, Shanghai Jiao Tong University, Shanghai 200240, China \\
$^2$Shanghai Research Center for Quantum Sciences, Shanghai 201315, China \\
$^3$Collaborative Innovation Center of Light Manipulation and Applications, Shandong Normal University, Jinan 250358, China \\
$^4$Department of Mechanical Engineering, and Institute for Physical Science and Technology, University of Maryland, College Park, Maryland 20742, USA, USA\\
$^{\dagger} $avikdutt@umd.edu, $^{\ast} $yuanluqi@sjtu.edu.cn}

\begin{abstract} 
The study of dissipative systems has attracted great attention, 
as dissipation engineering has become an important candidate towards manipulating light in classical and quantum ways. 
Here, we investigate the behavior of a topological system with purely dissipative couplings in a synthetic time-frequency space. 
An imaginary bandstructure is shown, where eigen-modes experience different eigen-dissipation rates during the evolution of the system, 
resulting in mode competition between edge states and bulk modes. 
We show that distributions associated with edge states can dominate over bulk modes with stable amplification once the pump and saturation mechanisms are taken into consideration, 
which therefore points to a laser-like behavior for edge states
robust against disorders.
This work provides a scheme for manipulating multiple degrees of freedom of light by dissipation engineering, 
and also proposes a great candidate for topological lasers with dissipative photonics.
\end{abstract}

\maketitle

\section*{Introduction}

Dissipation naturally exists in many physical systems and hence attracts broad interest. 
Through dissipation engineering, physical states of systems can be manipulated \cite{verstraete2009quantum, liu2011experimental, zippilli2021dissipative} in fields 
of ultra-cold atoms \cite{diehl2008quantum, diehl2010dissipation, seetharam2022correlation}, 
superconducting circuits \cite{mirrahimi2014dynamically, cohen2014dissipation, siddiqi2021engineering}, and 
photonics \cite{kippenberg2018dissipative, wanjura2020topological, wright2020mechanisms}. 
On the other hand, topological photonics shows non-trivial one-way edge states robust against imperfections \cite{hatsugai1993chern, wang2009observation, hasan2010colloquium, qi2011topological, ozawa2019topological, price2022roadmap}.
Therefore, combination of dissipation engineering and topology brings new physical phenomena
and provides potential applications in controlling quantum or classical states,
such as directional amplifiers {\cite{wanjura2020topological}}, quantum frequency locking \cite{nathan2020quantum} and quantum computation \cite{fujii2014measurement}.

Recent research on dissipative physics explores systems with complex couplings in lattice models in
the real space \cite{diehl2008quantum, diehl2010dissipation, fujii2014measurement, metelmann2015nonreciprocal}. However, when the number of lattice sites or the dimension of the system increases, the problem associated with spatial 
complexity becomes inevitable. Synthetic dimensions 
\cite{celi2014synthetic, yuan2018synthetic, lustig2021topological}, however, provide
alternative methods by utilizing other degrees of freedom of the system to reduce
the spatial complexity. By connecting discrete modes, 
artificial lattices with desired complex couplings can be constructed in synthetic dimensions \cite{boada2012quantum, regensburger2012parity, luo2015quantum, yuan2016photonic, ozawa2016synthetic, lustig2019photonic, hu2020realization}, 
providing a convenient way for studying topological physics with large-scale Hamiltonians or in high-dimensional
systems \cite{zhang2001four, price2015four, lian2016five, lohse2018exploring, yuan2018synthetic2, yu2021topological, wang2021generating, wang2021topological, li2023direct}. 
Moreover, synthetic dimensions bring exotic opportunities for manipulating different properties of light \cite{zhang2017manipulating, zhou2017dynamically, bell2017spectral, qin2018spectrum, dutt2020single, weidemann2020topological, li2022single, li2022observation}.
Recently, topological dissipation in a synthetic time dimension
has been studied by creating a time-multiplexed resonator network \cite{leefmans2022topological}.

In this paper, we study a dissipative synthetic two-dimensional (2D) time-frequency lattice model in a modulated resonator 
with multiple distinct circulating pulses. Dissipative couplings are introduced through 
auxiliary delay lines to connect pulses at different arrival times \cite{leefmans2022topological}. 
We use amplitude modulator (AM) to induce complex-valued connectivities 
between discrete frequency modes of pulses to construct the synthetic frequency dimension. 
A synthetic 2D imaginary quantum Hall model is then built and its bandstructure exhibits topological dissipation
with imaginary eigenvalues. 
We find that the field distribution initially localized at boundaries associated to edge states 
dominates initially but may eventually disappear because bulk bands hold larger negative imaginary eigenvalues (gain). 
However, by introducing saturation and external pump source, 
we obtain laser-like behavior for topological edge states, 
robust against disorders, which points towards topological lasers with synthetic dimensions.

\section*{Model}
\begin{figure}[htbp]
\centering
\includegraphics[width=0.9\textwidth ]{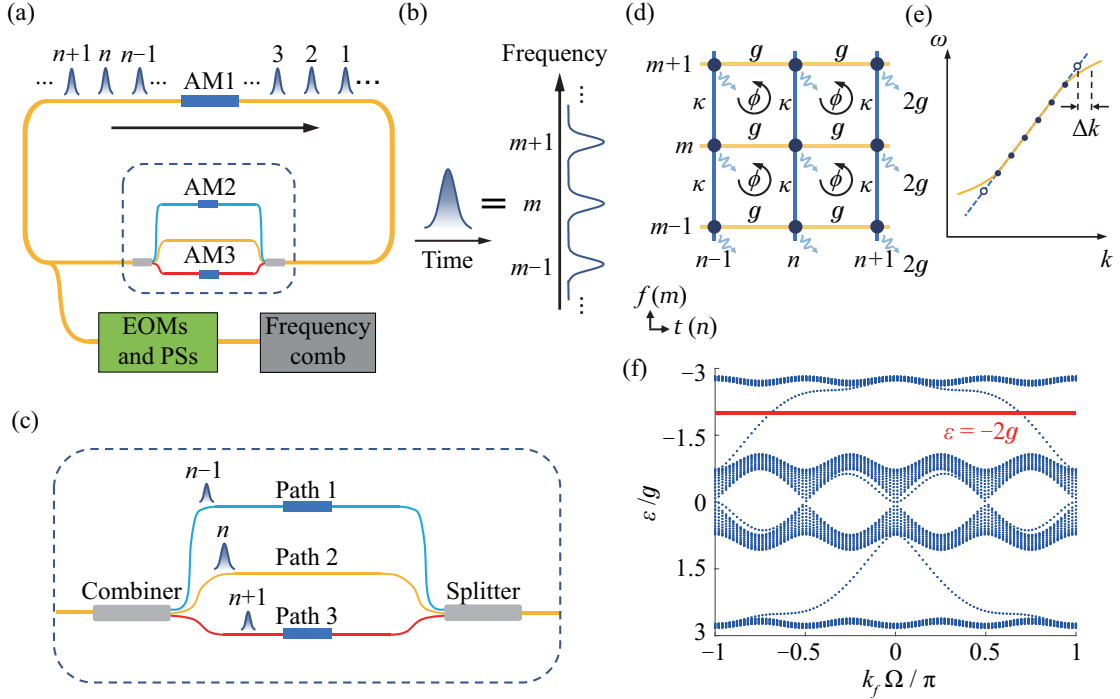}
\caption{ (a) Scheme of the cavity loop. The initial pulse chain and the external pumps are prepared by modulating strengths and phases of all frequency components of each frequency comb signal in real time by electro-optic modulators (EOMs) and phase shifters (PSs).
(b) Each pulse may carry multiple frequency modes, separated by $\Omega $. 
(c) Zoom-in of part including delay lines.
(d) A dissipative lattice in 2D synthetic time-frequency space. 
(e) Dispersion of waveguide that composes the cavity (yellow line). The dashed line represents a linear dispersion. 
Solid dots denote frequency modes, and hollow dots represent modes out of linear dispersion regime, 
introducing wavevector mismatch $\Delta k$. 
(f) Projected bandstructure with $\kappa = g$ and $\phi =\pi /2$.
Red line denotes $\varepsilon = -2g$.}\label{figure.1}
\end{figure}

We study a chain of $N$ pulses, separated with time interval $T$, circulating
inside a cavity loop in Fig.~\ref{figure.1}(a). AM is placed in the main cavity, described by a modulation function 
$\Gamma =1-2\alpha \cos[\Omega(t-z\cdot n_g/c)]$. Here $\alpha \ll 1$ is the modulation strength, $\Omega=h\cdot 2\pi/T_r $ is the modulation frequency (with $h$ being a positive real number), 
$T_r$ is the roundtrip time, $c$ is the speed of light in vacuum, 
and $n_g$ is the corresponding refractive index for the loop, 
which is considered to be the same for all frequency components in each pulse by assuming zero group velocity dispersion around the reference frequency $\omega _0$. 
Similar frequency modulation \cite{song2020two, yuan2021synthetic, li2022single} creates a synthetic lattice by connecting discrete frequency components at $\omega_m=\omega_0+m\Omega $, with $m$ being an integer.
Given $N$ and $T_r$, each pulse has a temporal width $<T_r/N$. 
We require $h \gg N$ so that all frequency components in each pulse are well separated in the frequency axis and hence each pulse holds a frequency-comb-like spectrum.

The cavity loop includes a pair of delay lines [see Fig.~\ref{figure.1}(c)], which provides dissipative couplings in time dimension \cite{leefmans2022topological}.
In detail, for the $n$-th pulse passing the splitter, a small portion of the pulse leaks into path 1 (3). 
The length of path 1 (3) is longer (shorter) than that of path 2
by $\Delta L=cT/n_g$. 
Therefore, the pulse that propagates through path 1 (3) is delayed (advanced) by $T$, i.e., encountering its nearby pulse  at
the combiner, which hence forms the synthetic lattice in time. 
Note that here, each pulse carries multiple frequency components separated by $\Omega$ as shown in Fig.~\ref{figure.1}(b), 
so compared with that passing path 2, 
each frequency component at $\omega _m$ passing path 1 or 3 accumulates an additional phase 
$\pm \left(\phi_{0}+m \phi\right)=\pm k_{m} \Delta L=\pm \left(\omega_{0}+m \Omega\right) n_{g} \Delta L / c$,
where $k_m$ is the wavevector, $\phi _0 \equiv \omega _0n_g\Delta L/c$, and $\phi \equiv \Omega n_g\Delta L/c$. 
This proposed system supports a 2D lattice in the time-frequency synthetic space, described by Hamiltonian (see Supplemental Material):
\begin{eqnarray}\label{6}
    \tilde{H}=-\sum_{m, n} i g\left[b_{m, n+1}^{\dagger} b_{m, n}+b_{m, n}^{\dagger} b_{m, n+1}\right]-\sum_{m, n} i \kappa\left[e^{-i n \phi} b_{m+1, n}^{\dagger} b_{m, n}+e^{i n \phi} b_{m, n}^{\dagger} b_{m+1, n}\right],
\end{eqnarray}
where $b_{m,n}$ and $b_{m,n}^{\dagger}$ are annihilation and creation operators for the $m$-th frequency mode of the $n$-th pulse, $g$ is the coupling amplitude between nearby pulses, and $\kappa = \alpha/T_r$.  Such system has an intrinsic global loss at $\varepsilon =-2g$ (see Supplemental Material).

The anti-Hermitian Hamiltonian (\ref{6}) gives 2D dissipative lattice structure in synthetic time-frequency space 
[see Fig.~\ref{figure.1}(d)]. 
We place AM2 (AM3) in path 1 (3) to switch off couplings in time dimension and create a temporal boundary. 
For the frequency dimension, one
can create artificial boundaries by choosing the dispersion of the waveguide that composes the cavity, as shown in Fig.~\ref{figure.1}(e), 
so only frequency modes hold resonant couplings are considered \cite{yuan2016photonic}.

We first plot projected bandstructure onto the wavevector that is reciprocal to the frequency dimension, 
if we consider 20 pulses with infinite frequency modes in Eq. (\ref{6}), with $\kappa = g$ and $\phi =\pi /2$
[see Fig.~\ref{figure.1}(f)].
$\tilde{H}$ possesses exclusively imaginary eigenvalues, suggesting each eigen-mode experiences 
different dissipation rate $\varepsilon $, instead of usual eigenvalues associated with eigen-frequencies in a Hermitian Hamiltonian. 
Here, a negative $\varepsilon $ represents gain, while a positive one represents loss. 
Note that there is an intrinsic global loss of $2g$ in the system (see Supplemental Material), so only eigen-modes with
$\varepsilon < -2g$ may have gain. One notes $\tilde{H} = iH_c$, where $H_c$ is a conservative Hamiltonian supporting a non-zero effective magnetic flux in 2D synthetic space in the current case \cite{fang2012realizing}.
Therefore, topological invariants of bands in Fig.~\ref{figure.1}(f) are identical to the ones for $H_c$, since the spectrum of $\tilde{H}$ is the same as $H_c$ except for the additional $i$ for the eigenvalues,
indicating that our system supports topological edge states, but in a dissipative way. 
We briefly discuss on the topological invariant of the system in the supplemental Material.

\begin{figure}[htbp]
 \centering
 \includegraphics[width=0.9\textwidth ]{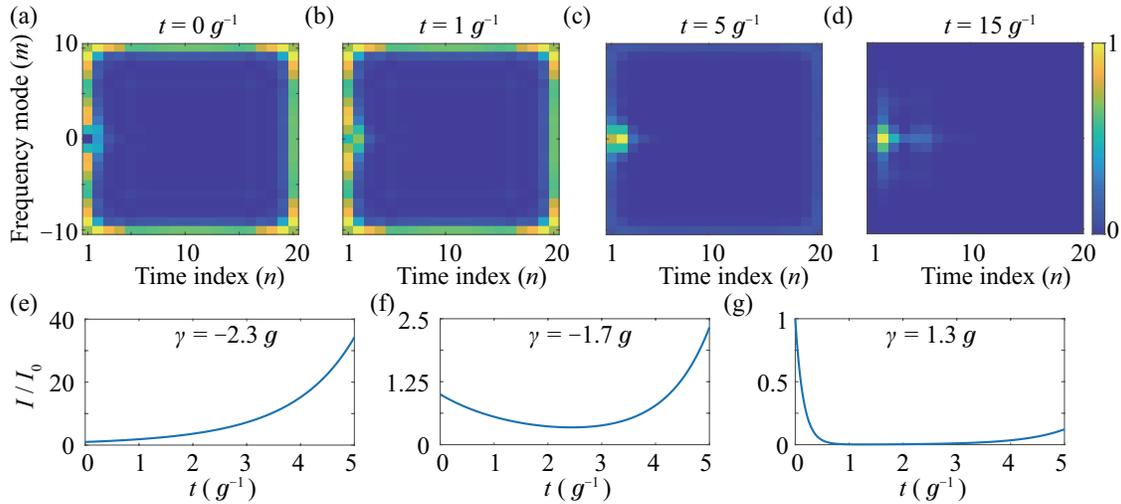}
 \caption{ (a) Normalized $|v_{m,n}|^2$ with $\gamma = -2.3$. Distributions of $|v_{m,n}|^2$ at different times, with (b) $t = 1g^{-1}$, (c) $t = 5g^{-1}$, and (d) $t = 15g^{-1}$. 
 Evolutions of $I$ for (e) $\gamma = -2.3g$, (f) $\gamma = -1.7g$, and
 (g) $\gamma = 1.3g$, respectively.}\label{figure.2}
\end{figure}

We further study a finite lattice by considering 20 pulses $(n \in [1,20])$ circulating inside the loop, 
each of which includes 21 resonant modes $(m \in [-10,10])$.
The evolution equation is,
\begin{eqnarray}\label{7}
{\frac{d\left|\Psi\right\rangle}{d t}}=-(i \tilde{H}+2g)\left|\Psi\right\rangle+{S}.
\end{eqnarray}
Here, $\left|\Psi\right\rangle = \sum_{m,n} v_{m,n} (t) b_{m,n}^\dagger |0\rangle$, where $v_{m,n} (t)$ is the field amplitude at the $m$-th mode in the $n$-th pulse, and ${S}$ denotes external pump source. If we assume $v_{m,n} (t) = \tilde v_{m,n}e^{-(\gamma+2g) t}$, where $\gamma $ represents the expected dissipation rate, and consider only the site (0,1), i.e., the 0-th mode at the 1-st pulse, is externally pumped at the strength $p$,
we obtain \cite{ozawa2014anomalous},
\begin{eqnarray}\label{8}
 g\left(\tilde v_{m, n+1}+\tilde v_{m, n-1}\right)+\kappa\left(e^{-i n \phi} \tilde v_{m-1, n}+e^{i n \phi} \tilde v_{m+1, n}\right)-\gamma\tilde v_{m, n}=p\delta_{m, 0} \delta_{n, 1}.
\end{eqnarray}
We set $p = \kappa = g$, choose $\gamma=-2.3g$ to excite edge states in Fig.~\ref{figure.1}(f), and diagonalize Eq. (\ref{8}) 
to obtain the initial distribution of $v_{m,n}(0) = \tilde v_{m,n}$ as shown in Fig.~\ref{figure.2}(a). 
Such choice of $\gamma$ results in the energy of the system distributed at boundaries of the synthetic space. 
Now we apply this initial distribution, 
which can be prepared by injecting all pulses with desired frequency distributions following $\tilde v_{m,n}$, and solve Eq. (\ref{7}) without external pump. 
Distributions of normalized $\left\lvert v_{m,n} \right\rvert^2$ at different $t$ are plotted in Fig.~\ref{figure.2}(b)-(d). 
Intensity distributions gradually leak into the bulk sites (a clearer bulk feature is provided with longer $t$, see Supplemental Material), which fundamentally differs from the case of a corresponding conservative system. 
The reason is that not
only the desired edge states at $\varepsilon =-2.3g$, but also a small portion of other states including bulk modes 
with larger negative dissipation rates $\left\lvert \varepsilon \right\rvert \sim 2.7g$ are initially excited. Although the energy localized in edge states 
is dominant initially, it gradually transfers 
to bulk states that have larger negative $\varepsilon$, as a result of mode competitions.
In experiments, $\gamma$ can be tuned by amplifier inside the main loop. 
We also perform simulations with initial distributions of $v_{m,n}$ using $\gamma = -1.7g$ and $\gamma = 1.3g$, 
respectively, linked to edge states, and similar phenomena are observed (see Supplemental Material). 
However, evolutions of total intensity $I \equiv \sum_{m,n} \left\lvert v_{m,n} \right\rvert^2$ are different 
[see Fig.~\ref{figure.2}(e)-(g)]. Here $I_0$ denotes reference intensity associated to initial distribution in Fig.~\ref{figure.2}(a). For the case with $\gamma < -2g$, $I$ monotonously increases, while for cases with $\gamma > -2g$, $I$ firstly decreases and later increases
then when modes with larger negative $\varepsilon $ dominate eventually as a result of mode competitions (see the Supplemental Material for a simple example in explaining the underlying physics).

\section*{Result}
\begin{figure}[htbp]
 \centering
 \includegraphics[width=0.9\textwidth ]{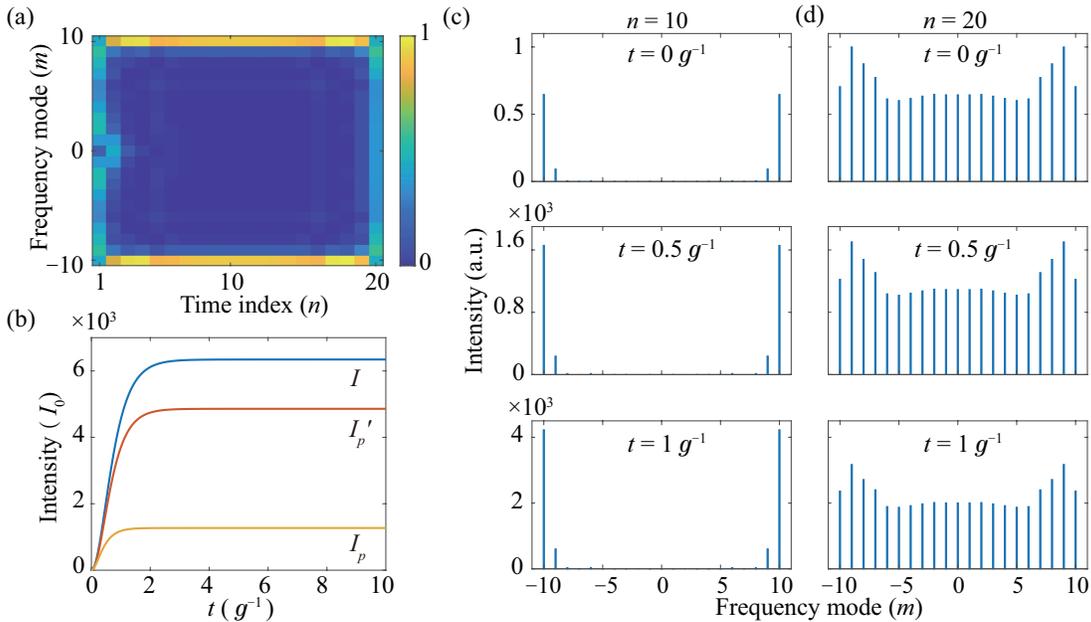}
 \caption{ (a) Intensity distribution of the steady state for simulating topological laser with Eq. (\ref{9}). 
 (b) Evolutions of $I$, $I_p{ }^{\prime}$, and $I_p$. 
 Intensity distributions for (c) 10-th pulse, and (d) 20-th pulse 
 at $t=0g^{-1}$, $t=0.5g^{-1}$ and $t=1g^{-1}$.}\label{figure.3}
\end{figure}

Although the dissipative topology here only exhibits features of topologically-protected edge states at small time which are eventually dominated by bulk modes with larger gains at longer time, 
the system still holds the capability for the realization of a topological laser. 
To this purpose, we add a saturation term into Eq. (\ref{7}) and evolutions of $v_{m, n}$ are:
\begin{eqnarray}\label{9}
 \dot{v}_{m, n}&=& -g\left(v_{m, n+1}+v_{m, n-1}\right)- \kappa\left(e^{-i n \phi} v_{m-1, n}+e^{i n \phi} v_{m+1, n}\right)-2 g v_{m, n}\nonumber\\
 &&+p_{m,n}-\frac{|\gamma| \sum_{m} \left\lvert v_{m,n} \right\rvert^2 }{2gI_s/(-\gamma-2g)+\sum_{m} \left\lvert v_{m,n} \right\rvert^2}.
\end{eqnarray}
The last term in Eq. (\ref{9}) describes saturation mechanism for the $n$-th pulse, 
which is dependent on the total intensity of all frequency components in each pulse, with
saturation intensity set as $I_s=25.7I_0$ \cite{yang2020mode}. $\gamma=-2.3$ to satisfy $-\gamma-2g>0$. 
We pump the system via $p_{m,n}$, which obeys the distribution in Fig.~\ref{figure.2}(a), 
i.e., we inject pulse sequences following $p_{m,n}$ into the system at every roundtrip. 
To achieve this, frequency combs can be used, with amplitudes and phases at each $\omega_m$ following $p_{m,n}$, and are injected into the loop at the $n$-th time slot [see Fig.~\ref{figure.1}(a)]. 
The system reaches a steady-state distribution shown in Fig.~\ref{figure.3}(a), which exhibits non-symmetric feature and is different from the eigen-state distribution in Fig.~\ref{figure.2}(a). 
The reason is that the saturation term for $v_{m,n}$ depends on summation of intensities on all frequency modes in the $n$-th pulse but edge states exhibit different distributions on frequency modes for the $1$-st, $20$-th pulses and the rest pulses. 
Therefore, one can see larger intensity distribution on the $\pm10$-th modes in pulses with $n \in[2,19]$ while nearly equal distributions on all modes in the 1-st and 20-th pulses. Figs.~\ref{figure.3}(c) and 3(d)
show intensity distributions in the 10-th and 20-th pulses at different times. 
To show the laser-like behavior, 
we also plot evolutions of total intensity $I = \sum_{m,n} |v_{m,n}|^2$ in Fig.~\ref{figure.3}(b). We compare $I$ with the pump intensity $I_p$ injected into the loop, which is calculated by solving Eq. (\ref{9}) with $\kappa=g=0$, i.e., no coupling terms in synthetic time-frequency dimensions. 
Furthermore, if we further ignore the loss term $-2gv_{m,n}$ and then solve Eq. (\ref{9}) with $\kappa=g=0$, we obtain the pure pump intensity $I_p{ }^{\prime}$.
One sees that $I$ increases faster than both $I_p$ and $I_p{ }^{\prime}$ initially and becomes saturated at $t\sim 2.4 g^{-1}$. 
We emphasize that the gain for the edge state mainly originates from the imaginary eigenvalue of the edge state supported by the system rather than the external pump (from the comparison between $I$ and $I_p$). 
The effect of the pump is to temporarily provide extra gain for edge states, keeping them dominant over bulk modes before saturation. 
Fig.~\ref{figure.3} implies 
that such topological dissipative photonics in the time-frequency space can be applied to a topological laser in synthetic dimensions \cite{yang2020mode}. 

Different from the topological laser which pumps the gain medium in a conservative topological system \cite{harari2018topological, bandres2018topological, yang2020mode}, we directly start with an exclusively dissipatively coupled system where the eigenstate with purely imaginary eigenvalue intrinsically has gain/loss. 
For a topological laser \cite{harari2018topological, bandres2018topological, yang2020mode}, gain originates from the external pump source since the system only supports real eigenvalues if the external pump is excluded.
The on-site gain provided by the external pump has no relevance to the eigenvalue of each mode. 
While in our model, gain/loss comes from the imaginary eigenvalues and thus edge modes and bulk modes have different gain/loss coefficients. 
Therefore, the implementation of the topological laser from the dissipative photonics is not straightforwad as bulk modes may have larger gains. 
We also compare effects of two simpler pump profiles (see Supplemental Material), which shows that the response of the system is affected by the pump profile, 
so the current choice of a complex pump profile leads to better gain performance.
This complex pump profile can be somehow simplified by only keeping most of the profile at boundaries in the synthetic space, which could be easier for the purpose of experiments (see Supplemental Material).

\begin{figure}[htbp]
\centering
\includegraphics[width=0.9\textwidth ]{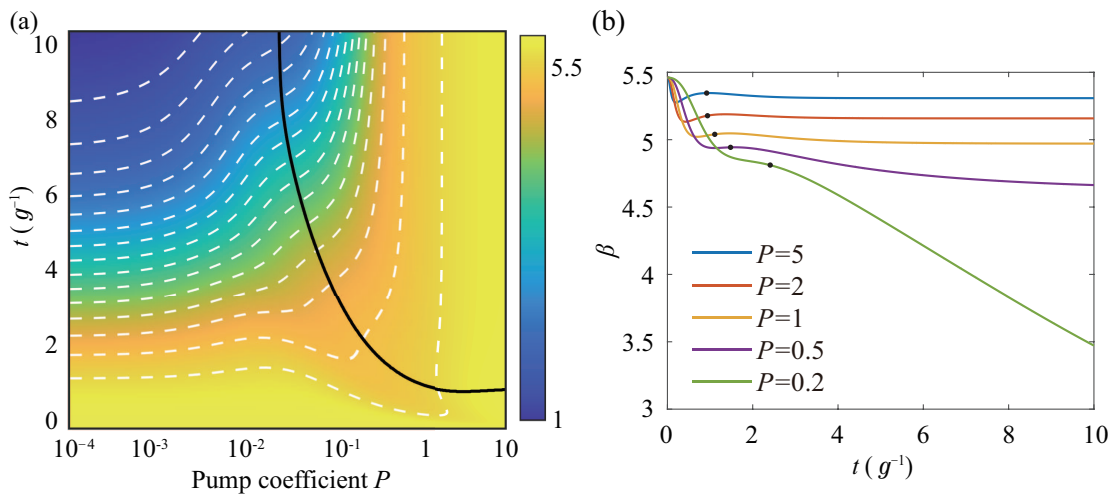}
\caption{ (a) $\beta $ versus time and $P$. 
White dashed lines denote equi-$\beta $ lines. The black solid line represents the saturation time. 
(b) $\beta $ versus time for different $P$. black dots represent the saturation time,  respectively.}\label{figure.4}
\end{figure}

We further investigate intensity distribution by varying the pump energy. 
In detail, we replace the pump term in Eq. (\ref{9}) by $P\cdot p_{m,n}$, where $P$ gives the pump coefficient that linearly changes amplitudes of pump pulses, 
so for $P=1$, the pump is the one used in Fig.~\ref{figure.3}.
We study distributions of edge and bulk modes during evolutions with different $P$ in Fig.~\ref{figure.4}(a), by defining the edge-bulk ratio $\beta =I_e/I_b$,
where $I_e\equiv \sum_{(m,n)} \left\lvert v_{m,n}\right\rvert^2$ with $(m,n)$ referring to all sites at boundaries in the synthetic space and $I_b\equiv I-I_e$.
One can see that for small $P$, 
the system initially holds a larger $\beta $, where most of the energy is localized at modes around boundaries. 
However, with the time evolution, $\beta $ decreases and the bulk modes are excited. 
On the other hand, for large $P$, large $\beta $ exists for a long time, 
indicating the persistence of edge modes during evolution. 
The saturation time, defined as the time when the increasing slope of the total intensity drops to half of its maximum, [see black line in Fig.~\ref{figure.4}(a)] 
and characterizing the time that the system reaches saturation, decreases when $P$ is increasing. 
To further understand such properties, we can classify $P$ into two regimes, i.e., $P\leqslant 0.01$ for the weak pump regime and $P>0.1$ for the strong pump regime.
For the strong pump regime, the system gets saturated within the time $ t\leqslant 10g^{-1}$. 
However, $\beta $ may still drop for smaller $P$. In Fig.~\ref{figure.4}(b), 
we plot $\beta$ versus $t$ for several choices of $P$. 
One sees that $\beta $ keeps at high ratio ($>4.5$) for $P\geqslant 0.5$, 
but decreases for the case of $P=0.2$ (see the Supplementary Material for details). 

\begin{figure}[htbp]
    \centering
    \includegraphics[width=0.9\textwidth ]{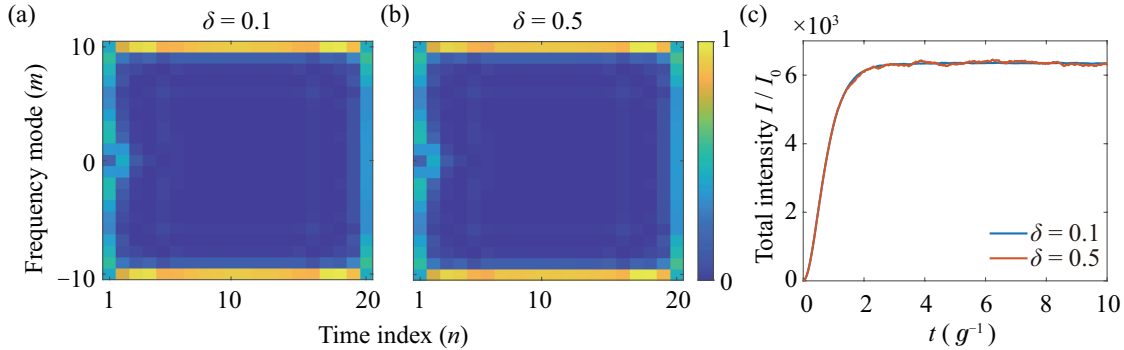}
    \caption{ Intensity distribution of the steady state with disorders included, where (a) $\delta =0.1$ (b) and $\delta =0.5$. 
    (c) Evolutions of $I$ versus time.}\label{figure.5}
\end{figure}

The features in the proposed synthetic time-frequency space exhibit dissipation but still hold topological properties. 
To demonstrate the topological protection, we introduce disorders in couplings terms. 
In particular, for modulations between frequency modes in the $n$-th pulse, $\kappa \rightarrow \kappa (1+\delta \cdot R_{1,n}) $, 
while for hoppings from the $n$-th pulse to $(n\pm 1)$-th pulse, $g \rightarrow g(1+\delta \cdot R_{2(3),n})$, 
where $\delta$ denotes the disorder strength, and $R_{1(2,3),n}$ are random numbers taken from $[-0.5,0.5]$. 
We perform simulations same as those in Figs.~\ref{figure.3}(a) and 3(b) but including disorders with $\delta =0.1$ and 0.5, respectively (see results in Fig.~\ref{figure.5}). 
We find that both steady-state distributions and intensity evolutions exhibit similar characteristics as those in Figs.~\ref{figure.3}(a) and 3(b), 
showing that the effect of disorder is negligible, and the proposed system, though dissipative, still exhibits topological protection. 

\begin{figure}[htbp]
    \centering
    \includegraphics[width=0.9\textwidth ]{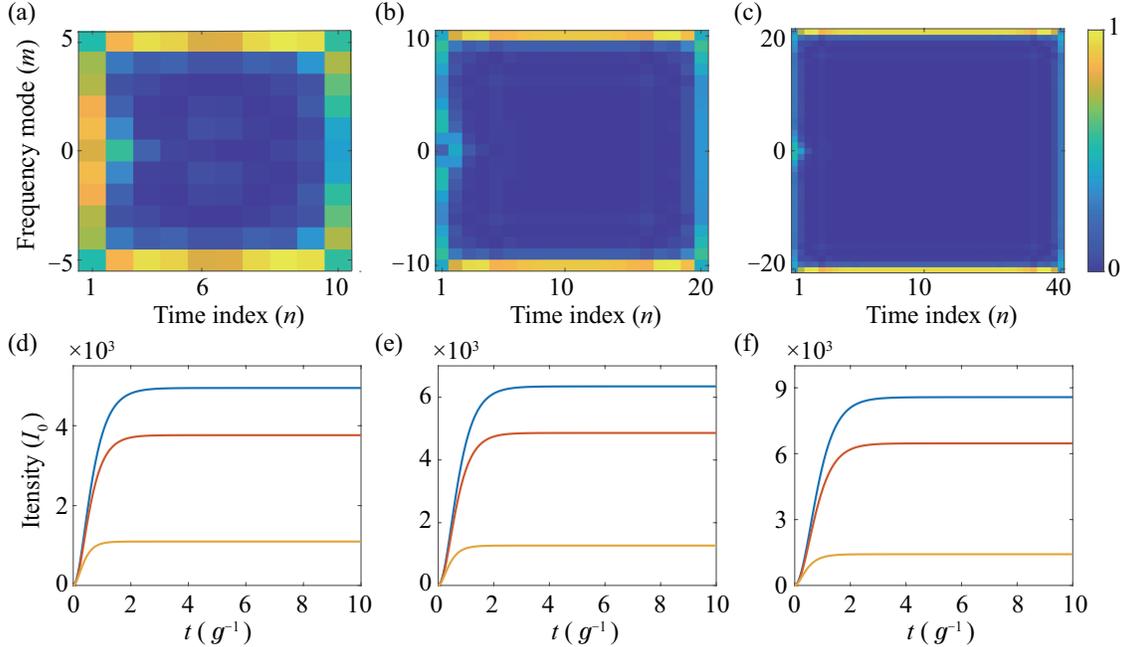}
    \caption{ Intensity distributions of the steady state for models with the lattice size as (a) $11\times 10$, (b) $21\times 20$, and (c) $41\times 40$ lattice size, respectively. 
    (d)-(f) Corresponding evolutions of $I$ (blue line), $I_p{ }^{\prime}$ (red line) 
        and $I_p$ (yellow line) versus time.}\label{figure.6}
\end{figure}

Before we end this section, we present the study of the size effects on the proposed system in simulations. 
We compare the steady state distributions and evolutions of the total intensity $I$ for models with the lattice size as $11\times 10$,  $21\times 20$, and $41\times 40$ respectively, which are shown in Fig. \ref{figure.6}. The saturation intensity is set as $I_s=25.7I_0$, where $I_0$ varies for the cases of different lattice sizes. The initial states and the pump profiles are obtained following the way in the previous section with $\gamma =-2.3g$ and pump coefficient $P=1$. 
One can find that for a smaller lattice size, more energy penetrates into the bulk sites, indicating that the topological protection is weaker. 
Moreover, the intensity distribution for $n=1$ is larger than the one for $n=10$ due to the intensity distribution of the initial state and the pump profile we select. 
However, one can see that if the larger lattice size is chosen,  the total intensity $I$ becomes larger. 
Moreover, as it can be seen in Fig. \ref{figure.6}(c), the distribution of the edge state becomes much more clearer, which shows that a system with larger size has better topological protection for edge states. In particular, for pulses with $n \in [2,39]$, one can clearly see the amplification of modes at edge, i.e., $m=20$ and $m=-20$. Such results may have an advantage in utilizing the lasing property of the edge state in the time-frequency space in a more efficient way.

\section*{Discussion}
Our proposal is experimentally feasible in a fiber network \cite{leefmans2022topological}. 
We emphasize that the separation of modes in both time and frequency dimensions should be larger than 
their corresponding full width at half maximum (FWHM) \cite{li2022single}. 
This requires that FWHM in time for each pulse is smaller than $T$ and FWHM in frequency for each mode is smaller than $\Omega$. 
For example, for a loop supporting $T_r \sim$ 200 ns, if there are $N=20$ pulses circulating, each pulse can have a temporal width $\sim 1$ ns to ensure that the pulses are separated in time. 
This then requires a modulation $\sim$ 10 GHz in frequency so that each pulse supports discretely spaced spectral components \cite{rueda2019resonant}. 
Furthermore, recent developments of on-chip integrated photonic technologies could also provide another possible experimental platform \cite{wang2018ultrahigh, balvcytis2022synthetic}. Limitations of our proposal may come from the additional losses from the propagation inside the loop, connections between components, etc. and also the challenge in synchronizing signal and pump pulses.

In summary, we propose a way to generate artificial lattices in a 2D synthetic time-frequency space and explore physical phenomena associated with dissipative photonics. 
The system supports imaginary eigenvalues, which results in gain for edge states. 
We study mode competition phenomena between edge states and bulk bands. 
By introducing saturation, we explore laser-like amplification with the topological protection, and find a way to excite such a lasing edge mode and preserve its dominance with the topological protection. The major difference between our work and Ref. \cite{leefmans2022topological} is the strategy of realizing the dissipatively coupled system. 
We study an active model, which itself can support the eigenstate with purely imaginary eigenvalue intrinsically having gain. 
Moreover, our model builds the effective magnetic flux in a passive way and also supports larger synthetic lattice size in experiments with the same spatial scale. 
The use of AM not only connects the sites in the frequency dimension but also plays a crucial role as exchanging energy between the system and the external reservoir, which distinguishes our work from previous works \cite{fujii2014measurement, leefmans2022topological, parto2023non}. 
Moreover, the dissipative couplings in frequency dimension from AM also differ our model from non-Hermitian models based on either on-site gain/loss \cite{weimann2017topologically} or direction-dependent gain/loss coupling \cite{yokomizo2019non}.
Our work therefore offers new opportunities in manipulating multiple properties of light and points towards topological lasers with synthetic dimensions \cite{st2017lasing, parto2018edge, zhao2018topological, harari2018topological, bandres2018topological, yang2020mode}, which may have applications in spectrotemporally shaped lasing emission as well as synchronous amplifications of multiple temporal pulses including many frequency components. Moreover, the proposed system may be generalized to mimic a quantum spin-Hall phase \cite{kane2005quantum, sheng2006quantum, bernevig2006quantum} with suitable design, and also can provide a realistic approach for further studying non-Hermitian physics and understanding dissipative topological systems \cite{shen2018topological, yokomizo2019non, song2019non, longhi2019probing, ashida2020non}.

\vspace{1cm}

\textbf{Acknowledgements}
The research was supported by National Natural Science Foundation
of China (12122407, 11974245, and 12192252), National Key Research and Development Program of China (No. 2021YFA1400900). L.Y. thanks the sponsorship from Yangyang Development Fund and the support from the Program
for Professor of Special Appointment (Eastern Scholar) at
Shanghai Institutions of Higher Learning. 
A.D. acknowledges support through the University of Maryland startup grant, 
and through grants from Northrop Grumman University Research Program and the National Quantum Lab (Q-Lab by UMD and IonQ).


\bibliographystyle{apsrev}

\bibliography{ref}

\end{document}